\documentclass{article}
\usepackage{arxiv}
\usepackage[utf8]{inputenc} 
\usepackage[T1]{fontenc}    
\usepackage{hyperref}
\usepackage{ upgreek }
\usepackage{url}            
\usepackage{booktabs}       
\usepackage{amsfonts}       
\usepackage{nicefrac}       
\usepackage{microtype}      
\usepackage{lipsum}
\usepackage{float}
\usepackage{graphicx}
\usepackage{multirow}
\usepackage{array}
\usepackage{comment}
\usepackage{longtable}
\usepackage{ragged2e}
\usepackage{subfig}
\usepackage{amsmath}

\title{Design and Development of an Autonomous Surface Vehicle for Water Quality Monitoring}

\author{
  MM Rashid   \\
  Mechatronics Engineering\\
  International Islamic University Malaysia\\
  Kula Lumpur, Malaysia\\
  \texttt{mahbub96@gmail.com} \\
   \And
      Rupal Roy \\
  Mechatronics Engineering\\
  International Islamic University Malaysia\\
  Kula Lumpur, Malaysia\\
  \texttt{rupal.roy@live.iium.edu.my} \\
   \And
  Md Manjurul Ahsan \\
  Industrial and Systems Engineering\\
  University of Oklahoma\\
  Norman, Oklahoma-73071 \\
  \texttt{ahsan@ou.edu} \\
   \And
 Zahed Siddique \\
  Department of Aerospace and Mechanical Engineering\\
  University of Oklahoma\\
  Norman, Oklahoma-73071\\
  \texttt{zsiddique@ou.edu}} 

\begin{document}
\maketitle

\begin{abstract}
Manually monitoring water quality is very exhausting and requires several hours of sampling and laboratory testing for a particular body of water. This article presents a solution to test water properties like electrical conductivity and pH with a remote-controlled floating vehicle that minimizes time intervals. An autonomous surface vehicle (ASV) has been designed mathematically and operated via MATLAB \& Simulink simulation where the Proportional integral derivative (PID) controller has been considered. A PVC model with Small waterplane area twin-hull (SWATH) technology is used to develop this vehicle. Manually collected data is compared to online sensors, suggesting a better solution for determining water properties such as dissolved oxygen (DO), biochemical oxygen demand (BOD), temperature, conductivity, total alkalinity, and bacteria.  Preliminary computational results show the promising result, as Sungai Pasu rivers tested water falls in the safe range of pH (~6.8-7.14) using the developed ASV.
\end{abstract}

\section*{Abbreviations}
ASV 	\quad	Autonomous Surface Vehicle\\
PID 		\quad Proportional Integral Derivative\\
SWATH	\quad Small Waterplane Area Twin-Hull\\ 
WQI	\quad	Water Quality Index\\

\section{Introduction}
\begin{figure}
    \centering
    \includegraphics{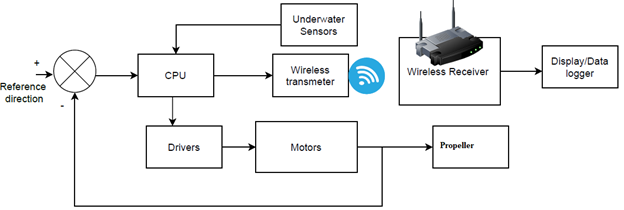}
    \caption{Block diagram of control system}
    \label{fig:fig1}
\end{figure}

The unmanned surface vessels (USVs) or the autonomous surface vehicles (ASVs) are crafts that are designed to allow autonomous control of marine platforms which are equipped with state-of-the-art sensors for carrying out a variety of missions like marine environment monitoring, hydrologic survey, target object searching, scientific study and so on~\cite{thompson2019review}. These vehicles are developed for shallow waters where it is difficult for survey vessels to reach and for areas dangerous to human operators like quarry lakes, contaminated waters~\cite{wang2009design}. Till date, many ASVs have been developed at academic labs, corporations, and governments for oceanic survey and defense purposes~\cite{vagale2021path}. Even though a large number of proven prototypes of ASVs have been designed and tested, there are only a few of them available in the industries for carrying out specified tasks. Especially when compared to the autonomous underwater vehicles, the ASV development area requires immense attention and further advancement~\cite{manley2008unmanned}.\\
The development of the first autonomous surface vehicles (ASV) dates back to World War II and used primarily for military purposes~\cite{corfield2002unmanned}. In the year 1946, ASVs were widely implemented in postwar applications, particularly to collect radioactive samples of water after the two atomic tests—Able and Baker— in the Pacific Ocean. In the 1950s-era, US Navy Mine Defense Laboratory's project DRONE constructed and tested a remotely operated minesweeping boat. By the 1960s remotely operated boats were used for missile firing purposes. The naval use of the unmanned surface vehicles has evolved over the years and today, the navy operates several USVs as target drones, including the Mobile Ship Target (MST), the QST-33, and QST-35/35A SEPTAR Targets, and the (HSMST) High-Speed Maneuverable Seaborne Target~\cite{bertram2008unmanned}.\\
One of the first Autonomous surface vehicles (ASVs) named ARTEMIS was developed in 1993 at the MIT Sea Grant Program~\cite{hinostroza2019cooperative}. Designed as a replica of a fishing trawler, its main use was testing navigation and control systems of the ASVs and was equipped with an electric motor and a servo actuated ruder. Here Proportional-plus derivative (PD) control system (Fig.~\ref{fig:fig1}) was used for simple heading control. Although ARTEMIS worked well as a test platform, it is not suitable for research in the coastal zone and the open ocean due to its small size and heading control. However, though ARTEMIS had limitations it gives a better concept about ASC~\cite{caccia2006autonomous}. 
A very important achievement in the field of autonomous surface vehicle development was the designing of the four surface crafts known as SCOUT (Surface Craft for Oceanographic and Undersea Testing) built by the MIT Department of Oceanic Engineering. It was a single hull kayak system design which makes it simple, robust, versatile, and had improved operational utility. The main concern of the SCOUT developers was low cost and high flexibility. Striking features about the SCOUT were that it could be controlled autonomously by preprogramming its controller; moreover, it could be controlled by an external computer via Wi-Fi or even via a computer that could give instructions directly to the propulsion and steering systems. In addition, Kayak design is popular due to its convenient mounting and loading~\cite{curcio2005experiments}. There are catamarans (twin hulls)~\cite{naeem2008design} and trimarans (triple hulls)~\cite{peng2009adaptive} system ASV, which are often chosen because of their higher system stability, payload capacity,  redundancy, and fewer risks of overturning in coastal water~\cite{campbell2012review}\\
Autonomous underwater vehicles (AUV) are compact, battery-powered submersibles equipped with their computers and capable of performing a wide range of tasks autonomously underwater~\cite{armstrong2019underwater}. AUVs have only been used in a semi-autonomous mode until recently, where they are constantly tracked and monitored~\cite{desa2006potential}. There are ongoing research efforts to provide real-time localization of multiple AUVs by using ASVs. The ASVs equipped with underwater acoustic modems, GPS, Wi-Fi, and 802.11b wireless Ethernet which used to achieve moving baseline data and help in testing adaptive behavior algorithms. The ASVs have successfully helped in acquiring position estimation and have increased the effectiveness of the AUC experiments~\cite{curcio2005experiments}.
The DELFIM ASC was developed to automatically record marine data and as a communication relay for a submarine and an auxiliary ship.  It facilitates the transfer of sonar and acoustic data via a vertically designed acoustic communication route.(Pascoal et al., 2000). The INFANTE is a self-contained AUV capable of operating at a maximum depth of 500 meters. The vehicle is a significant redesign of the MARIUS AUV, which was developed under the EU's MAST-I and MAST-I1 programs. Through the integration of motion sensors, the INFANTE AUV features an innovative navigation system~\cite{silvestre2004control}. \\
ASIMOV project is an advanced technological system where DELFIM ASC and INFANTE AUV two robotic ocean vehicles, were used designed, and built by the Institute for Systems and Robotics of the IST. It was developed for an acoustic communication link with the AUVs and with other underwater vessels using the ASVs. To serve this purpose a 3.5m catamaran with an indigenous wing-shaped central structure was developed. It was equipped with acoustic transducers, two-bladed propellers powered by electrical motors, and Doppler velocimeter. Normal ASV’s are low in speed but catamaran-style vehicles are more stable and faster for their monohull designs~\cite{caccia2006autonomous}.\\
Water is a natural resource that must be preserved, protected, and monitored continuously as it is the basic elements, necessary for the survival of not only humans but all kinds of life forms on earth~\cite{inyinbor2018water}. Water quality monitoring is highly essential for tracking pollutants, the presence of harmful elements, and in turn, ensuring clean non-toxic water supply for humans as well as aquatic life~\cite{rao2013design}. The various parameters that used to monitor the water quality includes Dissolved Oxygen (DO), Biochemical Oxygen Demand (BOD5), Chemical Oxygen Demand (COD), Suspended Solids (SS), Ammoniacal Nitrogen (AN), pH and so on. WQI (Water Quality Index) is used to classify the monitored river water quality. The WQI is based on the Interim National Water Quality Index (INWQS), which is defined as a set of standards based on beneficial uses of water. To classify the river water quality that is monitored there is a Water Quality Index. A Water Quality Index (WQI) assigns quality value to a set of rough data that is collected from water samples. The main purpose of WQI is to interpret the collected data and make it simple  for people to understand,  whether the water quality is good or bad~\cite{zainudin2010benchmarking}.\\
HydroNet ASV is a small-sized ASV, designed for monitoring the coastal water quality which detects heavy metal concentrations such as chrome: Cr (VI), mercury: Hg (II), and cadmium: Cd (II) from water in real-time using custom-made miniaturized onboard sensors. The USV is designed for long-range missions, lodging an onboard water analysis system. It was tested in the Livorno coastal area distance of 12 682.6 m~\cite{ferri2011design,ferri2014hydronet}. Subsequently, the HydroNet modification was used to collect water samples up to 50 m from the water column. The system was retested in field trials at Leverkusen, but they do not perform any chemical analyzes of the water, which was left for future work~\cite{fornai2016autonomous}. Springer vehicle is another USV designed as a double-hole vessel with a medium seaplane and can be used in a variety of ways as a mission water profile and payload. It was developed for environmental monitoring and pollutant tracking. An integrator sensor was used for navigation purposes~\cite{naeem2008design}\\
At present low-resolution, water quality monitoring is being carried out by various laboratories~\cite{aardema2019high}. However, there are many disadvantages of the manual approach~\cite{chen2018water}. The collection of data by this method is not continuous in space and time and thus, occasionally occurring pollutants are missed. Also, this method is time-consuming and expensive as it involves frequent visits by an individual to specific. Specific water properties such as dissolved oxygen content and oxidation-reduction potential are susceptible to change in pressure and altitude, and therefore these parameters must be measured on-spot for better accuracy. Laboratory testing approach provides slow, inaccurate and sparse results which are difficult to be interpreted and tabulated~\cite{rao2013design}\\
The autonomous surface crafts are becoming exceedingly popular for water quality monitoring especially in coastal and estuarine waters because of their ability to provide real-time on- spot data without the need of taking the water samples to the laboratory for testing and data generation~\cite{ubina2022review}. These unmanned vehicles equipped with in-situ instrumentation are capable of collecting data from places inaccessible to the human population like water with high algal concentration. These unmatched advantages of collecting spot data with real-time transmission and reaching inaccessible areas have led to the increased popularity of ASVs in the field of water quality surveillance~\cite{steimle2006unmanned}.
Every day along with the coastal areas, harsh winds, swell waves, and other phenomena control sand transport and change ocean morphology thus playing a vital role in the dynamic behavior of coastal regions. This constantly changing ocean morphology needs continuous monitoring which is achieved through Bathymetric surveys that generate maps for the surveyed areas. These maps illustrate ocean bed topography and help in creating vessel navigation charts~\cite{ferreira2009autonomous}. The ever-growing demand for bathymetric surveys and the inaccessibility of some coastal areas to the human population have created the demand for autonomous surface crafts~\cite{shojaei2018proof}. Various autonomous surface crafts like the solar-powered OASIS platform that investigated the nature and extent of harmful algal blooms, the Wave Glider powered by wave hydel energy and the ASV developed at ETH Zurich with a winch mechanism that measures varying oceanic depths (up to 130m deep)~\cite{valada2014development}.\\
From the above discussion, we can see that many of these surface crafts discussed are exceptionally high cost and developed on a very large scale thus restricting their usage to large scale industrial projects. Also, many of the surface crafts are semi-autonomous and have short battery life~\cite{wolfe2021evaluation}. Thus, there is a need for the development of an autonomous surface craft that can operate in riverine and estuarine environments and provide cost-effective and portable long-duration data gathering.  
Wireless sensor networks (WSNs) were developing swiftly withinside the beyond few years~\cite{bhushan2020requirements}. Lots of studies has been carried out on WSN communications, WSN strength conservation, WSN routing algorithms, etc. However, maximum studies particularly specialize in terrestrial sensor networks. Research on underwater sensor networks is constrained. Considering that constrained into account, in this study, water first-rate tracking systems have been accessed that can assist with water pollutants detection and discharge of poisonous chemical substances and infection in water. As a consequence, temperature, pH, dissolved oxygen, connectivity and turbidity parameters are monitored in river/lake water first-rate tracking systems.\\ The summary of the technical contribution can be express as follows:  
\begin{enumerate}
    \item Design a suitable mechanical structure for an autonomous surface vehicle, capable of floating and maneuvering on the surface of a water body;
    \item Develop a sensory mechanism for the ASV to measure water quality parameters on the spot;
    \item Develop navigation and control techniques for the surface vehicle to aid in monitoring water quality and collecting bathymetric data and; 
    \item Evaluate the performance of the autonomous surface vehicle
\end{enumerate}
The remainder of the section is organized as follows: Section~\ref{method} discussed materials and methods. Section~\ref{result} describes our study's findings and examines the general conclusions. Section~\ref{conclusions} concludes by summarizing the general findings.
\section{Materials and methods}\label{method}
\subsection{Mechanical Design of ASV}
\begin{figure}
    \centering
    \includegraphics{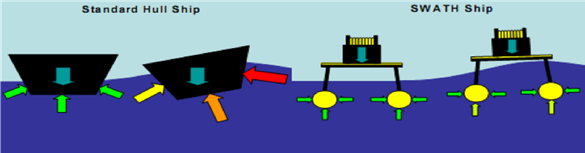}
    \caption{SWATH design gives stability to the vehicle}
    \label{fig:swath}
\end{figure}

\begin{figure}
     \centering
     \includegraphics{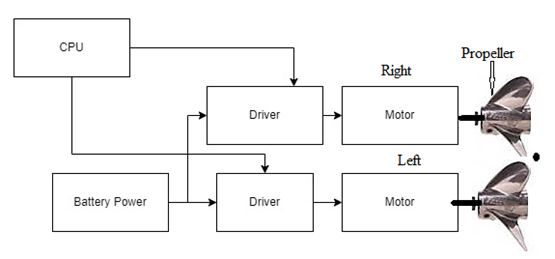}
     \caption{Block diagram of propulsion system}
     \label{fig:block}
 \end{figure}

The Small Waterplane Twin Hull Design (SWATH) design is chosen for this autonomous surface vessel (Fig.~\ref{fig:swath}). The design is selected to reduce roll and pitch disturbances which lead to errors in on-spot measurement. SWATH design gives better wave disturbance rejection and excellent stability during roll and pitch motions as it has less hull volume at the waterline, giving it lesser buoyancy change during wave interaction.   
The following Table~\ref{tab:tab1} and ~\ref{tab:tab2} shows the data collected from the tests:
\begin{table}
\caption{Dynamic analysis data of the SWATH style ship ‘wasp’~\cite{mahacek2005dynamic})}
    \centering
    \begin{tabular}{ccccccc}\toprule
          & Freeboard&	Cycles& \multicolumn{2}{c}{Displacement}	&	Time&	Period\\\midrule
         &in&	n&	B1&	    B(n+1)& $\Delta$ t(s)&	P\\\midrule 
        \multirow{3}{*}{Roll}&	16&	6&	   7&	3.5&	37.90&	6.32\\
	&12&	5&	10.5&	4.5&	36.33&	7.27\\
	&8	&6	&  13	&4.5	&41.60	&6.93\\
\multirow{3}{*}{Pitch
(Black)}	&16&	2&	1.5&	0.5&	12.61&	6.31\\
	&12&	4&	3.5&	0.75&	27.66&	6.91\\
	&8&	4&	3.25&	0.75&	25.74&	6.44\\
\multirow{3}{*}{Pitch
(Front)}&	16&	2&1.5&	0.5&	12.26&	6.13\\
	&12&	5&	2.5&	1&	34.46&	6.89\\
	&8&	6&	2.75&	1&	38.48&	6.41\\\bottomrule

    \end{tabular}
    
    \label{tab:tab1}
\end{table}
\begin{table}
\caption{Dynamic analysis data of the SWATH style ship ‘wasp’~\cite{mahacek2005dynamic}}
    \centering
    \begin{tabular}{p{.1\linewidth}p{.1\linewidth}p{.1\linewidth}p{.1\linewidth}p{.1\linewidth}p{.1\linewidth}p{.1\linewidth}}\toprule
         Logarithmic
decrement&	Damping ratio&	Damped frequency&	Natural
frequency&
	Stiffness&	Damping&	Critical damping
constant\\\midrule
$\delta$ (delta)&	$\zeta$ (zeta)&	$\omega$  d& $\omega$	n&	k (Ibs/ft)	&c(lbs-sec/ft)&	cc\\\midrule
0.116&	0.00919&	0.99459&	0.99464&	24.58&	0.454&	49.42\\
0.169&	0.01348&	0.86474&	0.86482&	19.74&	0.616&	45.66\\
0.177&	0.01407&	0.90619&	0. .90627&	22.96&	0.713&	50.66\\
0.549&	0.04367&	0.99638&	0.99733&	24.71&	2.164&	49.56\\
0.385&	0.03063&	0.90876&	0.90919&	21.82&	1.470&	48.00\\
0.367&	0.02916&	0.97641&	0.97682&	 26.67&	1.592	&54.61\\
0.549&	0.04367&	1.02482&	  1.02580&	 26.14&	 2.226&	50.97\\
0.183&	0.01458&	0.91166&	  0.91166&	 21.94&	 0.702&	48.14\\
0.169&	0.01342&	0.97960&	  0.97969&	 26.83&	 0.735&	     54.77\\\bottomrule

    \end{tabular}
    
    \label{tab:tab2}
\end{table}
To find the stability of the boat, the following linear second-order equation was used:
\begin{equation}
m\ddot{x}+c\dot{x}+kx=f(t)
\end{equation}
Where, ‘m’ is the mass of the system, ‘c’ is the damping, which is a factor of the horizontal surface area and slows vertical motion, and ‘k’ is the stiffness which is a factor of the change in displacement versus the change in height. For a typical system, high values would imply a significant resistance to input forces; however, with aquatic vessels, higher damping and, more specifically, high stiffness correspond to greater wave tracking. For the SWATH case, both values should be low compared to the mass, which they are. Thus, it can be predicted that the vessel will have a low response to waves for the targeted sea state.\\ 
By testing at different heights, the optimal loading and operating range can be determined. For each of the test settings, the middle data set (which corresponds to 12’’free board) gives the lowest stiffness. This is because it is using the thinner section of the struts, whereas the lower setting ended up using the reserve flotation, and the higher setting brings the pontoons out of the water in extreme cases. Using the equation 2, the system is defined as stable if both roots have negative and real parts, which occurs when m, c, and k have the same sign and is a rare possibility in the case of catamaran and kayak style boats.
\begin{equation}
    x=\frac{-b\pm\sqrt{b^2-4ac}}{2a}=s1,\ s2	
\end{equation}
For the SWATH scenario, all three are positive, indicating that the SWATH system is more stable here in this range.
After making the design choice, the surface vehicle ‘s main hull is made from various shapes and types of PVC pipes which are positioned together at 90° angles to form a structure, capable of staying afloat on water and carrying a payload of up to 5kgs. On top of the PVC, boat structure is mounted as a platform to house the electrical components that are the microcontroller and the sensors.\\   
The following components and materials are chosen to conduct the experiment:
\begin{itemize}
    \item The Hull: Hull is composed of Polyvinyl Chloride (PVC) pipes of various diameters. PVC pipes are ideal for building the boat structure as they are lightweight and can be easily installed. They can sustain low temperatures and are greatly resistant to corrosion. Being low cost, biological, and chemical resistant and having easy workability, it is being used in a wide variety of applications. 
\item	Propulsion System: The propulsion system is composed of four Seabotix BTD thrusters. The thrusters are brushless DC motors encased in a watertight housing. These thrusters help in forwarding movement and also aid in turning and maneuvering in narrow spaces (Fig.~\ref{fig:block}). 
 Platform: A platform made of wood is mounted on top of the pipe arrangement completing the structure of the vehicle. It is used to house the microcontroller and other electric equipment used. A sturdy wooden platform was chosen as the platform had to be cut in several places for the installment of sensors and other connections. 
\item 	Electronic Component Enclosure: Electronic components form an integral part of the autonomous vehicle system and need to be given proper protection during the operation. A watertight plastic case is used for the placement of all the electrical equipment in the system. Two pipe casings adjacent to each thruster are provided for the Electrical Conductivity and pH sensor for water quality monitoring. 
\end{itemize}
\subsection{Navigation and Data Transmission system of ASV}
\begin{figure}
    \centering
    \includegraphics[width=\textwidth]{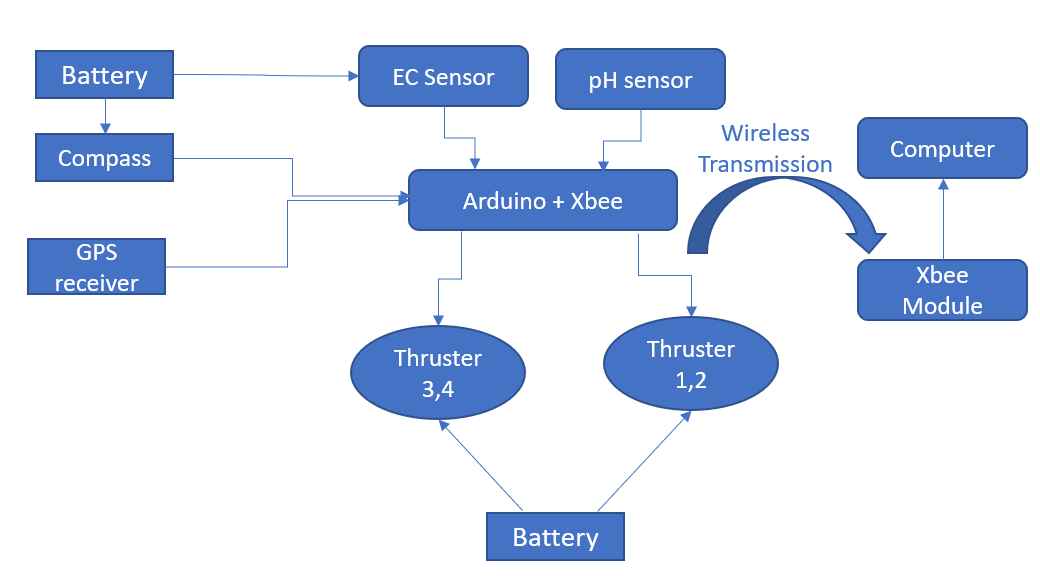}
    \caption{Electronics System Diagram of the Autonomous Surface Vehicle}
    \label{fig:EV}
\end{figure}

The propelling force that drives the vehicle comes from the four thrusters. The sensors (EC sensor, pH sensor, Depth Sensor) are the components that provide the data to be transmitted. The microcontroller, the GPS receiver, and the compass would be interfaced with the Arduino microcontroller which will perform all data processing. Programming in the Arduino environment is much easier as compared to the conventional C/C++ programming. Fig.~\ref{fig:EV} demonstrates the electronics systems diagram used to develop the autonomous vehicle.
Since the boat is an autonomous surface vehicle, there is no tether between the boat at the water body and the computer on the shore. The Arduino microcontroller is preprogrammed the location coordinates from where the water samples need to be tested and it converts numerical values to pulse width modulation (PWM) signals for the motor controller and the boat moves. The sensors interfaced with the Arduino at the water body provide data to it and through the Xbee modules communicating wirelessly with each other, the data is automatically transferred from the water body onto the shore. Thus, transmission between the Xbee Shield and the Xbee connected to the computer onshore helps to transfer sensor readings in real-time.  
\subsection{Kinematic Analysis of the ASV }
The first step in modeling the vehicle is to choose the coordinate reference frames. Two coordinate frames are chosen for the autonomous vehicle. One is the fixed frame of reference (XYZ) such that at time t = 0, the CG (center of gravity) of the vehicle will be at the origin. The second frame chosen is the body-fixed frame (BFF) such that its origin lies at the CG of the vehicle as shown in the Fig.~\ref{fig:body} below: 
\begin{figure}
    \centering
    \includegraphics{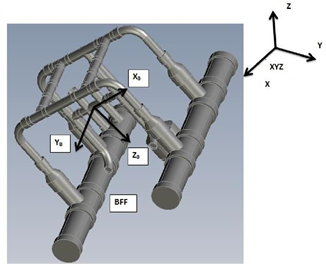}
    \caption{The body-fixed axis and the Global axis of the ASV}
    \label{fig:body}
\end{figure}
The second step is to assign the degrees of freedom to our vehicle. Following the popular assumption of neglecting the heave, pitch, and roll motions which does not have any effect on the accuracy of the vehicle while maneuvering to reach its target. Thus, the configuration vector of the fixed frame XYZ for the body frame BFF is given as: 
\begin{equation}
    \eta (t) = [X, Y, \uppsi], t \geq 0                          
\end{equation}
The surge (movement along $X_b$ axis) designated by u(t), sway (movement along $Y_b$ axis) designated by v(t), and yaw (rotation about the $Z_b$ axis) designated by r(t) are the three motions that are taken into consideration. The velocity vector is then represented as: 
\begin{equation}
         V(t) = [u, v, r], t \geq 0    
\end{equation}
Taking angle of trim ($\theta$) and angle of the roll ($\upgamma$) as negligible, the rotation matrix from the fixed to the body reference frame is given as: 
\begin{equation}
    J\left(\eta\right)=\left[\begin{matrix}cos\psi&-sin\psi&0\\sin\psi&cos\psi&0\\0&0&1\\\end{matrix}\right]
\end{equation}\\
From equations (3), (4) \& (5):
\begin{equation}
    \frac{d(\eta\left(t\right))}{dt}=\dot{\eta}\left(t\right)=J\left(\eta\right)V\left(t\right)
\end{equation}
Equation (6) depicts the relation between $\eta$ (t) and V(t) and these together represent the kinematic equations of the ASV. 
\subsection{Motion Analysis of the ASV}
The next step is to generate the ASV's equations of motion, which is accomplished by taking into account the external forces and torques acting on water. Taking the forces and moments along the three directions of the fixed body frame, equations can be developed as follows: 
For Surge (movement along the $X_b$ axis), 
\begin{equation}
X_b = m(\dot{u} -vr - yg\dot{r} - x_g r^2 )
\end{equation}
For Sway (movement along the $Y_b$ axis), \begin{equation}
    Y_b = m (  \dot{v}  - ur - xg\dot{r}  - ygr^2 )
\end{equation}
For Yaw (rotation about the $Z_b$ axis), 
\begin{equation}
    N = Iz \dot{r} +m [xg(\dot{r}+ur)-yg (\dot{u}- vr)]
\end{equation}
$X_b$, $Y_b$, N are the forces and the moment acting on the vehicle and $x_g$ and $y_g$ are the distances from the origin of the fixed body reference to the CG of the vehicle. When $x_g$ and $y_g$ $\rightarrow 0$, the simplified matrix form of the motion equations can be written as: 
\begin{equation}
    M\dot{V}(t)+[C(v) +D(v) ]V(t)+g[\eta(t)] = T_R (t) ;  t  \geq 0
\end{equation}
Where M represents the mass- inertia matrix. $C [ v(t) ] \epsilon  R^{3\times3}$ represents the coriolis and centripetal matrix. $D [v(t)]  \epsilon R^3$ represents the damping vector, $g [\eta(t)]  \epsilon R^3$ is the gravitational force and moments vector. Finally $T_R (t) = [X_b(t), Y_b(t), N(t)]^T$ is the input vector of the forces and moments. The mass- inertia and Coriolis matrix involve the rigid body mass and added mass which occur due to the force of the fluid in contact with the vehicle. Thus, matric M and C [v(t)] can be re-written as:  
\begin{equation}
    M = M_{RB} + M_{AM}                                                   
\end{equation}

where,
\begin{center}
 $M_{RB}=\left[\begin{matrix}m&0&0\\0&m&0\\0&0&Iz\\\end{matrix}\right]$\\
$C_{RB}=\left[\begin{matrix}0&-mr&0\\mr&0&0\\0&0&0\\\end{matrix}\right]$   
\end{center}

Since we are assuming the heave motion to be negligible in this case, the gravitational matrix $g [\eta (t)] = 0$.
\subsection{Dynamic Analysis of the ASV }
To develop a complete mathematical model for the ASV, it is important to consider the hydrodynamic forces that are the drag and the various kinds of frictional forces acting on the vehicle. First, the damping forces in the $X_b$ direction acting on the hull can be given as: 
\begin{equation}
    F_h,X_b(u,r) = C_f \rho A_h U \mid U\mid
\end{equation}
This force on the hull is due to the viscous drag and depends on Cf, the frictional resistance coefficient, $\rho$, the density of water in  $m^3$, wetted area of the hull, $A_h$ and the U, the distance from CG to the port side of the hull in the $Y_b$ direction. Now, integrating the damping forces for both hulls to get the total damping force in the $Y_b$ direction: 
\begin{equation}
    F_h, Y_b (v,r) = \int_{-dax}^{dfx}{1/2CDhTh\rho v(x)\mid v(x)\mid dx}
\end{equation}
Here,
\begin{center}
$v\left(x\right)=v+\ \frac{vbow-vstern}{Lh}$
\end{center}
Having completed the damping force analysis, the next step is to analyze the thrust force that acts on the vehicle. A propulsion system aids the movement of the vehicle to the desired location at a particular velocity and acceleration. It is the propellers that generate thrust.
T that drives the vehicle in water which is given by:
\begin{equation}
    T = K_T\rho   d^4 n^2
\end{equation}
The torque of the shaft attached to the propeller is represented by the equation: 
\begin{equation}
    Q = K_Q\rho   d^5 n^2                                                                
\end{equation}
The input forces and the moments put together to give us the thrust vector $T_R$:
\begin{equation}
    T_R= [ X Y N]
\end{equation}
Let $T_P$ and $T_S$ be the thrusts delivered by the port side and star side of the vehicle respectively and $\alpha_p$ and $\alpha_S$ be the inclinations to the $X_b$ axis. Also, $d_1$ and $d_2$ are taken as distanced in $X_b$ and $Y_b$ directions. Thus, the thrust vector is given by: 
\begin{equation}
   X= T_P cos\alpha_p + T_s cos\alpha_S                                            
\end{equation}
\begin{equation}
 Y= T_P sin\alpha_P- T_s sin\alpha_S                                                  
\end{equation}
\begin{equation}
N= X.d_1 + Y.d_2                                                           
\end{equation}
From eq (8) we have:
\begin{equation}
   \dot{V} (t) = - M^{-1}[ C(v) + D(v) ] V(t) + M^{-1}TR 
\end{equation}
Defining state vector: 
\begin{equation}
   X = [ \eta^T     V^T ]^T                                                            
\end{equation}
From (6) and (18) State Space Representation of the ASV is given as: 
\begin{equation}
    \dot{\mathbf{X}} = f (X) + BT_R
\end{equation}
\begin{equation}
    B =\left[\begin{matrix}0\\M^{-1}\\\end{matrix}\right] 
\end{equation}
\begin{equation}
        f (X) = \begin{bmatrix}
        J(\eta)V(t)\\
        -M^-1[C(v) +D(v)]V(t)\\
        \end{bmatrix}
\end{equation}
\subsection{Nomoto Model for Heading Control of the ASV}
\begin{figure}
    \centering
    \includegraphics{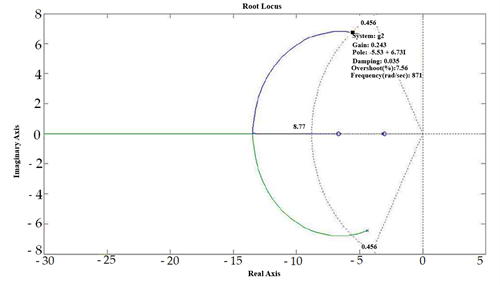}
    \caption{Root Locus of the stable $2^{nd}$ Order Nomoto Model}
    \label{fig:root}
\end{figure}

After the kinematic and dynamic analysis of the ASV, the next step is to develop a suitable controller for its heading control. For this purpose, a Nomoto Model is chosen which is a ship steering model utilizing first order second-order models for ship autopilot design. The Nomoto Model is highly popular in steering autonomous vehicles as it is simple and accurate, which is highly essential considering the nonlinear and complex hydrodynamics involved with boat steering.\\
The transfer function of the second order Nomoto Model is given by: 
\begin{equation}
    \frac{r(s)}{\delta(s)}=\frac{Kr(1+T3)}{(1+T1s)(1+T2s)}
\end{equation}
where $K_r$ is the yaw rate gain, and $T_1$, $T_2$, and $T_3$ are time constants. The numerical values for these constants are adapted. After substituting the numerical value of the constants, the transfer function obtained is given as:
\begin{equation}
    \frac{r(s)}{\delta(s)}=\frac{1.9103s+5.799}{0.3037s^2+0.7488s-1}
\end{equation}
Using the second-order Nomoto Model and plotting the root locus and step response showed that the system was unstable as one of the poles fell on the right-hand plane. To address this issue, a PID controller was utilized in conjunction with the Nomoto Model. After plotting the root locus, step response, and bode plots, it was observed that the system became stable due to the adoption of the PID controller as shown in Fig.~\ref{fig:root}.
Therefore, the Nomoto Model is used for the heading control of the autonomous surface vehicle in case the vehicle is thrown off track due to wind, heavy rain, and other disturbances. The controller helps the vehicle to follow its desired course irrespective of the physical disturbances by receiving feedback from the compass.
\section{Result and Discussions}\label{result}
\begin{figure}
    \centering
    \includegraphics{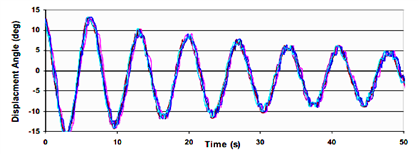}
    \caption{Side-to-side response of the ASV with 8” of freeboard (distance from the waterline to the boat top)}
    \label{fig:fig7}
\end{figure}

\begin{figure}
    \centering
    \includegraphics{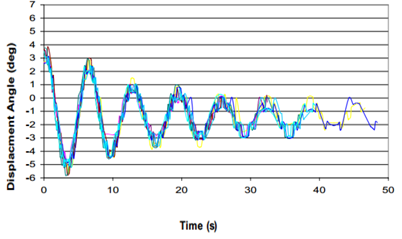}
    \caption{Side-to-side response of the bow of the ASV with 8” of freeboard}
    \label{fig:fig8}
\end{figure}

\begin{figure}
    \centering
    \includegraphics{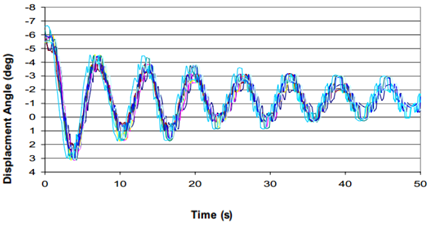}
    \caption{Side-to-side response of the stern of the ASV with 8” of freeboard}
    \label{fig:fig9}
\end{figure}

\begin{figure}
    \centering
    \includegraphics{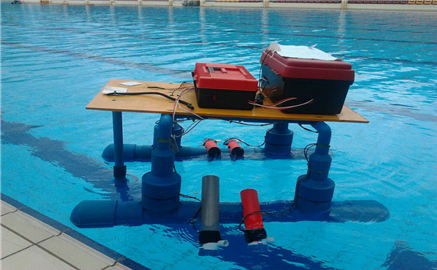}
    \caption{ASV measuring Water Quality Parameters}
    \label{fig:fig10}
\end{figure}

\begin{figure}
    \centering
    \includegraphics{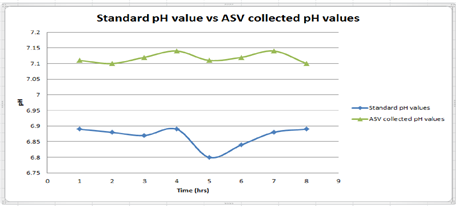}
    \caption{Comparative analysis of pH results}
    \label{fig:fig11}
\end{figure}

The first step in determining the performance of the ASV was to perform its motion analysis using the SolidWorks software. This motion analysis gives an idea of the stability of the ASV on the surface of the water which is the most important feature of a waterborne vehicle. The following graphs (Fig.~\ref{fig:fig7}) depict the angular displacement (in terms of pitch and roll) of the ASV with time.  The slow decrease in a response implies low damping and stiffness which is characteristic of a SWATH style boat as mentioned earlier. Fig.~\ref{fig:fig7} shows that within 50 seconds, the side-to-side response of the boat decreases, and it gains stability. 
The separate responses of the bow (front side) and the stern (rear side) to the waves are also obtained depicting individual stability as shown in Fig.~\ref{fig:fig8} and \ref{fig:fig9} respectively.
After analyzing the movement of the boat in water, the next step is to collect data. For this purpose, water samples were collected from the Sungai Pasu river Malaysia and tested for pH and Electrical Conductivity at the Environmental Laboratory at the Biotechnology Engineering Department. Fig.~\ref{fig:fig10} displays the developed ASV that measure the quality parameters. 
Several readings were taken at different times and recorded. After that, the ASV was used to measure the pH and electrical conductivity of the river water. The sensor data was wirelessly transmitted from the boat to the computer simultaneously via email using the XCTU software. During the water monitoring, the compass sends its readings continuously to the station computer at the shore, giving the heading of the ASV. This reading is in degrees as measured from the North. The compass is placed on the ASV such that the stern (front) of the boat points to north, thus any deviation of the stern from its path is recorded and transmitted by the compass.\\
After testing the ASV at the water, a comparison was made between the standard pH value vs transmitted pH value (as shown in Fig.~\ref{fig:fig11}) and the standard EC value vs transmitted EC value (as shown in Fig. 12) of the ASV. data collected at the lab using standard method and values of pH and EC obtained and transmitted by the ASV.  Fig. 9 depicts that pH values obtained from the lab are very close to the values obtained by the ASV. However, there was a disparity of about 1-2 mS /cm (millisiemens per centimeter) between the Electrical Conductivity (EC) data collected by the two methods. The main reason for this was that data in the lab was collected at a low temperature of about 21◦C and hence the EC of water was lower as compared to the values collected by the ASV at an outside temperature of around 31◦C. Note that, Electrical Conductivity is temperature-dependent (increasing approximately 2-3\% per degree Celsius).
\begin{figure}
    \centering
    \includegraphics{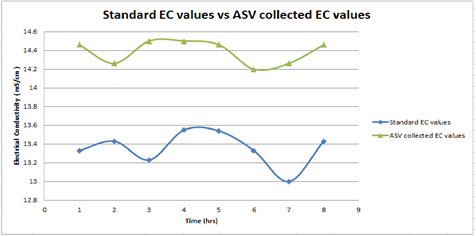}
    \caption{Comparative analysis of Conductivity results}
    \label{fig:fig12}
\end{figure}
The Table~\ref{tab:tab3} below shows the Malaysian Water Quality Index, which assigns an acceptable range to the various water quality parameters. The purpose of a WQI is to summarize large amounts of water quality data for a specific river into simple terms. This makes it easily understandable for communities in water quality monitoring. The acceptable pH for various classes of water depending on their use are in the range of 6.5 to 9 and thus, it can be said that Sungai Pasu River water tested by the ASV falls in the safe range of pH (~6.8-7.14). \\
However, the Electrical Conductivity of the river water (13-14mS/cm), when compared to the Water Quality Index, is far above the acceptable range of 1-6 mS/cm. High Electrical Conductivity indicates high dissolved-solids concentration; dissolved solids can affect the suitability of water for domestic, industrial, and agricultural uses and can even be hazardous to aquatic life. Additionally, high dissolved-solids concentration can cause deterioration of plumbing fixtures and appliances~\cite{zainudin2010benchmarking}.
\begin{table}
\caption{3 Malaysian Water Quality Index~\cite{zainudin2010benchmarking}}
    \centering
    \begin{tabular}{cccccccc}\toprule
         Parameters&	Unit&\multicolumn{6}{c}{	Classes}  \\\midrule
         Ammoniacal Nitrogen&	mg/l&	0.1&	0.3&	0.3&	0.9&	2.7&	>2.7\\
         BOD&	mg/l&	1&	3&	3&	6&	12&	>12\\
COD&	mg/l&	10&	25&	25&	50&	100&	>100\\
DO&	mg/l&	7&	5-7&	5-7&	3-5&	<3&	<1\\
pH&	mg/l&	6.5-8.5&	6.5-9.0&	6.5-9.0&	5-9&	5-9&	-\\
Color&	TUC&	15&	150&	150&	-&	-\\	
Elec. Conductivity&	$\mu$ S/cm&	1000&	1000&	-&	-&	6000&	-\\\bottomrule

    \end{tabular}
    
    \label{tab:tab3}
\end{table}
From the data collected by the ASV, followed by its analysis, it is clear that the IIUM river needs water-treatment processes, such as coagulation, flocculation, sedimentation, and disinfection, to remove excessive dissolved solids from it and thus improve its quality.  
\section{Conclusions}\label{conclusions}
In this work, an autonomous surface vehicle (ASV) was designed and tested to carry out water quality monitoring. The kinematic and dynamic modeling of the autonomous vehicle was done using the universal motion and moment equations. Also, a heading controller was developed to keep the vehicle on its course during harsh winds and other disturbances. The SWATH style structure of the vehicle gave it more stability while carrying out missions on the water surface. The use of Smart Drive motor drivers for driving the thrusters helped each thruster to move independently thus making differential steering possible for the vehicle. The main controller of the vehicle is the Arduino Mega 2560 which gave instructions to the thrusters to move in the desired direction. The navigation system of the vehicle which consists of a GPS receiver and compass was interfaced with the Arduino and wireless communication between the ASV and control center was made possible by the use of Xbee modules.  The developed ASV can be used to collect field data for water quality assessment through wireless remote sensing. In addition, it can be used for field measurements in inaccessible and toxic waters. Another important contribution of ASV could be the rapid measurement of water quality after natural disasters such as floods and hurricanes. Future works include but are not limited to ASV models integrated with machine learning techniques~\cite{ahsan2021effect,roy2021review} and artificial intelligent-based ASD data analysis.
\bibliographystyle{unsrt}  
\bibliography{main}

\begin{thebibliography}{10}

\bibitem{thompson2019review}
Fletcher Thompson and Damien Guihen.
\newblock Review of mission planning for autonomous marine vehicle fleets.
\newblock {\em Journal of Field Robotics}, 36(2):333--354, 2019.

\bibitem{wang2009design}
Jianhua Wang, Wei Gu, and Jianxin Zhu.
\newblock Design of an autonomous surface vehicle used for marine environment
  monitoring.
\newblock In {\em 2009 International Conference on Advanced Computer Control},
  pages 405--409. IEEE, 2009.

\bibitem{vagale2021path}
Anete Vagale, Rachid Oucheikh, Robin~T Bye, Ottar~L Osen, and Thor~I Fossen.
\newblock Path planning and collision avoidance for autonomous surface vehicles
  i: a review.
\newblock {\em Journal of Marine Science and Technology}, pages 1--15, 2021.

\bibitem{manley2008unmanned}
Justin~E Manley.
\newblock Unmanned surface vehicles, 15 years of development.
\newblock In {\em OCEANS 2008}, pages 1--4. Ieee, 2008.

\bibitem{corfield2002unmanned}
SJ~Corfield.
\newblock Unmanned surface vehicles and other things.
\newblock In {\em Proceedings of the Unmanned Underwater Vehicle Showcase 2002
  Conference, Southampton, UK}, pages 83--91, 2002.

\bibitem{bertram2008unmanned}
Volker Bertram.
\newblock Unmanned surface vehicles-a survey.
\newblock {\em Skibsteknisk Selskab, Copenhagen, Denmark}, 1:1--14, 2008.

\bibitem{hinostroza2019cooperative}
MA~Hinostroza, Haitong Xu, and C~Guedes Soares.
\newblock Cooperative operation of autonomous surface vehicles for maintaining
  formation in complex marine environment.
\newblock {\em Ocean Engineering}, 183:132--154, 2019.

\bibitem{caccia2006autonomous}
Massimo Caccia.
\newblock Autonomous surface craft: prototypes and basic research issues.
\newblock In {\em 2006 14th Mediterranean Conference on Control and
  Automation}, pages 1--6. IEEE, 2006.

\bibitem{curcio2005experiments}
Joseph Curcio, John Leonard, Jerome Vaganay, Andrew Patrikalakis, Alexander
  Bahr, David Battle, Henrik Schmidt, and Matthew Grund.
\newblock Experiments in moving baseline navigation using autonomous surface
  craft.
\newblock In {\em Proceedings of OCEANS 2005 MTS/IEEE}, pages 730--735. IEEE,
  2005.

\bibitem{naeem2008design}
W~Naeem, T~Xu, R~Sutton, and A~Tiano.
\newblock The design of a navigation, guidance, and control system for an
  unmanned surface vehicle for environmental monitoring.
\newblock {\em Proceedings of the Institution of Mechanical Engineers, Part M:
  Journal of Engineering for the Maritime Environment}, 222(2):67--79, 2008.

\bibitem{peng2009adaptive}
Yan Peng, Jian-da Han, and Qing-jiu Huang.
\newblock Adaptive ukf based tracking control for unmanned trimaran vehicles.
\newblock {\em International Journal of Innovative Computing, Information and
  Control}, 5(10):3505--3516, 2009.

\bibitem{campbell2012review}
Sable Campbell, Wasif Naeem, and George~W Irwin.
\newblock A review on improving the autonomy of unmanned surface vehicles
  through intelligent collision avoidance manoeuvres.
\newblock {\em Annual Reviews in Control}, 36(2):267--283, 2012.

\bibitem{armstrong2019underwater}
Roy~A Armstrong, Oscar Pizarro, and Christopher Roman.
\newblock Underwater robotic technology for imaging mesophotic coral
  ecosystems.
\newblock In {\em Mesophotic Coral Ecosystems}, pages 973--988. Springer, 2019.

\bibitem{desa2006potential}
Elgar Desa, R~Madhan, and P~Maurya.
\newblock Potential of autonomous underwater vehicles as new generation ocean
  data platforms.
\newblock {\em Current science}, pages 1202--1209, 2006.

\bibitem{silvestre2004control}
Carlos Silvestre and A~Pascoal.
\newblock Control of the infante auv using gain scheduled static output
  feedback.
\newblock {\em Control Engineering Practice}, 12(12):1501--1509, 2004.

\bibitem{inyinbor2018water}
A~Inyinbor~Adejumoke, O~Adebesin~Babatunde, P~Oluyori~Abimbola, A~Adelani
  Akande~Tabitha, O~Dada~Adewumi, A~Oreofe~Toyin, et~al.
\newblock Water pollution: effects, prevention, and climatic impact.
\newblock {\em Water Challenges of an Urbanizing World}, 33:33--47, 2018.

\bibitem{rao2013design}
Aravinda~S Rao, Stephen Marshall, Jayavardhana Gubbi, Marimuthu Palaniswami,
  Richard Sinnott, and Vincent Pettigrovet.
\newblock Design of low-cost autonomous water quality monitoring system.
\newblock In {\em 2013 International Conference on Advances in Computing,
  Communications and Informatics (ICACCI)}, pages 14--19. IEEE, 2013.

\bibitem{zainudin2010benchmarking}
Zaki Zainudin.
\newblock Benchmarking river water quality in malaysia.
\newblock {\em Jurutera}, 12:15, 2010.

\bibitem{ferri2011design}
Gabriele Ferri, Alessandro Manzi, Francesco Fornai, Barbara Mazzolai, Cecilia
  Laschi, Francesco Ciuchi, and Paolo Dario.
\newblock Design, fabrication and first sea trials of a small-sized autonomous
  catamaran for heavy metals monitoring in coastal waters.
\newblock In {\em 2011 IEEE International Conference on Robotics and
  Automation}, pages 2406--2411. IEEE, 2011.

\bibitem{ferri2014hydronet}
Gabriele Ferri, Alessandro Manzi, Francesco Fornai, Francesco Ciuchi, and
  Cecilia Laschi.
\newblock The hydronet asv, a small-sized autonomous catamaran for real-time
  monitoring of water quality: From design to missions at sea.
\newblock {\em IEEE Journal of Oceanic Engineering}, 40(3):710--726, 2014.

\bibitem{fornai2016autonomous}
Francesco Fornai, Gabriele Ferri, Alessandro Manzi, Francesco Ciuchi, Francesco
  Bartaloni, and Cecilia Laschi.
\newblock An autonomous water monitoring and sampling system for small-sized
  asvs.
\newblock {\em IEEE Journal of Oceanic Engineering}, 42(1):5--12, 2016.

\bibitem{aardema2019high}
Hedy~M Aardema, Machteld Rijkeboer, Alain Lefebvre, Arnold Veen, and Jacco~C
  Kromkamp.
\newblock High-resolution underway measurements of phytoplankton photosynthesis
  and abundance as an innovative addition to water quality monitoring programs.
\newblock {\em Ocean Science}, 15(5):1267--1285, 2019.

\bibitem{chen2018water}
Yiheng Chen and Dawei Han.
\newblock Water quality monitoring in smart city: A pilot project.
\newblock {\em Automation in Construction}, 89:307--316, 2018.

\bibitem{ubina2022review}
Naomi~A Ubina and Shyi-Chyi Cheng.
\newblock A review of unmanned system technologies with its application to
  aquaculture farm monitoring and management.
\newblock {\em Drones}, 6(1):12, 2022.

\bibitem{steimle2006unmanned}
Eric~T Steimle and Michael~L Hall.
\newblock Unmanned surface vehicles as environmental monitoring and assessment
  tools.
\newblock In {\em OCEANS 2006}, pages 1--5. IEEE, 2006.

\bibitem{ferreira2009autonomous}
Hugo Ferreira, C~Almeida, A~Martins, J~Almeida, N~Dias, A~Dias, and E~Silva.
\newblock Autonomous bathymetry for risk assessment with roaz robotic surface
  vehicle.
\newblock In {\em Oceans 2009-Europe}, pages 1--6. Ieee, 2009.

\bibitem{shojaei2018proof}
Alireza Shojaei, Hashem~Izadi Moud, and Ian Flood.
\newblock Proof of concept for the use of small unmanned surface vehicle in
  built environment management.
\newblock In {\em Proceeding of Construction Research Congress}, pages
  148--157, 2018.

\bibitem{valada2014development}
Abhinav Valada, Prasanna Velagapudi, Balajee Kannan, Christopher Tomaszewski,
  George Kantor, and Paul Scerri.
\newblock Development of a low cost multi-robot autonomous marine surface
  platform.
\newblock In {\em Field and service robotics}, pages 643--658. Springer, 2014.

\bibitem{wolfe2021evaluation}
Jessica~Simmerman Wolfe.
\newblock {\em Evaluation of the utility and performance of an autonomous
  surface vehicle for mobile monitoring of waterborne biochemical agents}.
\newblock PhD thesis, Mississippi State University, 2021.

\bibitem{bhushan2020requirements}
Bharat Bhushan and Gadadhar Sahoo.
\newblock Requirements, protocols, and security challenges in wireless sensor
  networks: An industrial perspective.
\newblock In {\em Handbook of computer networks and cyber security}, pages
  683--713. Springer, 2020.

\bibitem{mahacek2005dynamic}
Paul Mahacek.
\newblock Dynamic analysis of a swath vessel.
\newblock {\em MBARI Internship Report}, pages 1--13, 2005.

\bibitem{ahsan2021effect}
Md~Manjurul Ahsan, MA~Mahmud, Pritom~Kumar Saha, Kishor~Datta Gupta, and Zahed
  Siddique.
\newblock Effect of data scaling methods on machine learning algorithms and
  model performance.
\newblock {\em Technologies}, 9(3):52, 2021.

\bibitem{roy2021review}
Rupal Roy, Maidul Islam, Nafiz Sadman, MA~Mahmud, Kishor~Datta Gupta, and
  Md~Manjurul Ahsan.
\newblock A review on comparative remarks, performance evaluation and
  improvement strategies of quadrotor controllers.
\newblock {\em Technologies}, 9(2):37, 2021.

\end{thebibliography}

\end{document}